\input{aipcheck}

\documentclass[
    ,final            
  ]
  {aipproc}

\layoutstyle{6x9}
\newcommand{\cd}{\makebox[0.08cm]{$\cdot$}}

\begin{document}

\title{Recent developments in light-front dynamics}

\author{V.A. Karmanov}{
  address={Lebedev Physical Institute, Leninsky Prospekt 53,
119991 Moscow, Russia}
}

\begin{abstract}
Recent results on relativistic few body systems, obtained in the framework
of light-front dynamics, are briefly reviewed. The following subjects are
discussed: two scalar bosons with ladder and cross ladder kernel; two
fermions with OBE kernel; relativistic scattering (elastic and inelastic);
three bosons and fermions with zero-range interaction; many-body
contributions.
\end{abstract}

\maketitle


\section{Introduction}
In relativity, the state vector of a composite system is defined, in
general, on a space-like surface. A limiting case the space-like plane --
light-front (LF) plane, defined by the equation $\omega\cd x=0$, with
$\omega^2=0$, is very preferable. In this case, the bare vacuum state (an
eigenstate of free Hamiltonian) is also an eigenstate of full Hamiltonian.
This simplifies the theory a lot. Dynamics, determining the LF wave
functions, is called light-front dynamics (LFD). If $\omega$ is a general
four-vector (but always $\omega^2=0$), we get explicitly covariant version
of LFD \cite{cdkm}. In particular case $\omega=(1,0,0,-1)$ we recover the
standard version \cite{bpp}.

LFD is successfully applied to relativistic few-body systems and to the
field theory \cite{cdkm,bpp}. Below we present some recent applications to
relativistic few-body systems.
\section{Two scalar bosons}
The orientation of LF plane $\omega\cd x=0$ is determined by the direction
of $\vec{\omega}$, i.e., by the unit vector
$\vec{n}=\vec{\omega}/|\vec{\omega}|$.  Wave functions, defined on the
LF plane, depend  on   $\vec{n}$. For a two-body wave function we get:
$$\psi=\psi(\vec{k},\vec{n}).$$ For two spinless constituents with zero
angular momentum wave function depends on the scalar products:
$\psi=\psi(\vec{k}\,^2,\vec{n}\cd\vec{k})$. For systems with non-zero
spins and total angular momentum the vector $\vec{n}$ participates in
construction of the angular momentum on the equal ground with the relative
momentum $\vec{k}$.

Equation for two-body wave function $\psi(\vec{k},\vec{n})$ is determined
by the kernel $V(\vec{k},\vec{k}\,',\vec{n},M^2)$ which also depends on
$\vec{n}$ and, in addition, on the two-body mass $M$. For a given model,
the kernel is calculated by the LFD graph technique \cite{cdkm,bpp}. For
the ladder exchange, its nonrelativistic (NR) counterpart is the Yukawa
potential $V(r)=-\alpha\exp(-\mu r)/r$, where $\alpha=g^2/(16\pi m^2)$
(for scalars) and $g$ is the coupling constant in the interaction
Hamiltonian $H^{int}=-g\psi^2\varphi$.
\begin{figure}[!ht]
\includegraphics[height=.27\textheight]{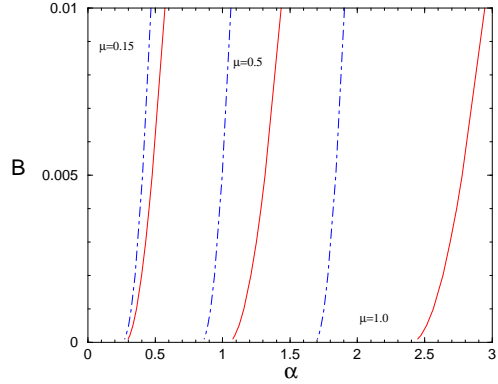}
\caption{Two-body binding energy $B$ versus coupling constant $\alpha$.
Solid curve is the LFD calculation. Dash-dotted curve is non-relativistic
result.} \label{fig1}
\end{figure}

The NR  and LFD calculations \cite{MJ} with the ladder kernel are compared
in figure \ref{fig1}. Calculation by the Bethe-Salpeter (BS) equation is
very close to the LFD one, which differs from NR. For heavy exchanged mass
$\mu$ (comparable with the constituent mass $m$) the nonrelativistic and
relativistic results strongly differ from each other even for very small
binding energy. This conclusion is important since $\rho$ and $\omega$
mesons, incorporated in the $NN$ potential, are heavy. Therefore, the
nonrelativitic approach  may be too approximate.

We calculated also \cite{ck04}, both by the BS and LFD equations, the
binding energy incorporating sum of ladder and cross-ladder graphs. The
stretched box graphs (with two non-crossed intermediate mesons) were also
taken into account, but their contribution  turned out to be small. The
result of calculation, for exchanged mass $\mu=0.5$, is shown in figure
\ref{fig4}. We see that the contribution of the crossed ladder graphs,
relative to the ladder exchange, is large. The difference between the LFD
and BS results is still small. The influence of the BS crossed box was
also analyzed in \cite{td01}.
\begin{figure}[!ht]
\includegraphics[height=.27\textheight]{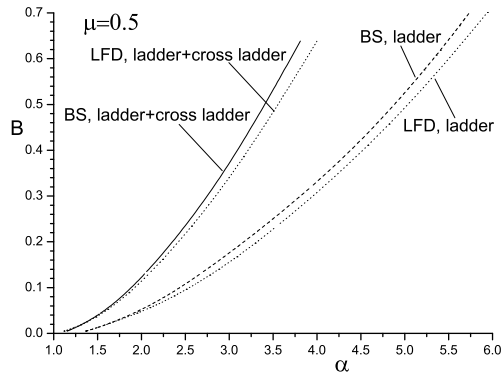}
\caption{Binding energy    $B$  vs. coupling constant $\alpha$,
incorporating ladder and  cross ladder contributions.}\label{fig4}
\label{fig4}
\end{figure}
\section{Two fermions}
LF wave function of two-fermions with the zero total angular momentum
$J=0^+$  has the following general form: $$ \psi(\vec{k},\vec{n})=
\frac{1}{\sqrt{2}} \left(f_1+ \frac{i\vec{\sigma}\cd [\vec{k}\times
\vec{n}]} {\sin\theta}f_2\right), $$ where $\theta$ is angle between
$\vec{k}$ and $\vec{n}$. It is determined by two spin components $f_1,f_2$
depending on $\vec{k}\,^2,\vec{n}\cd\vec{k}$ (or, equivalently, on $k$ and
$\theta$) and satisfying a system of two equations.

The wave function for $J=1^+$ is determined by six spin components.
Corresponding system of 6 equations is split in two uncoupled subsystems
of 4 and 2 equations.

Two fermion bound states with the OBE kernel incorporating scalar,
pseudoscalar and vector  mesons were investigated in detail, exchange by
exchange, in \cite{mck03}. The form factors in the $NN$-boson vertices
were introduced and stability relative to cutoff was studied. For the
scalar exchange~\cite{mck01} (Yukawa model), a stable bound state with
$J=0^+$ exists without form factor, if the coupling constant
$\alpha=g^2/(4\pi)$  does not exceed a critical value $\alpha_c=3.72$. If
$\alpha
>\alpha_c$, cutoff is needed to obtain a finite binding energy. For
pseudoscalar exchange the state $J=0^+$ is stable, whereas for the $J=1^+$
state a vertex form factor is required. It is always required for the
vector exchange.

Relativistic deuteron wave function (with six spin components) was
calculated in \cite{ck95} and applied to the deuteron e.m. form factors in
\cite{ck99}.  The experimental data \cite{t20} on $t_{20}$, later obtained
at JLab, are on the curves from \cite{ck99}.
\section{Relativistic scattering}
Two following important features of relativistic scattering equations
should be emphasized \cite{ji1,ji2,ock}. ({\it i})~Relativistic kernel
$V(\vec{k}\,',\vec{k},\vec{n},M^2)$ \emph{automatically} takes into
account inelasic channel. When  $M> 2m+\mu$,  then denominator in the
ladder kernel  may cross zero, that results in a singularity and, in its
turn, in imaginary part of the phase shift $\delta$. ({\it ii})~For the
exchange kernel, the amplitude is not unitary above threshold, i.e.,
$\delta$ does not satisfy the condition
$Im(\delta)=k^2\sigma^{inel}/(4\pi)$ (valid for small $Im(\delta)$). In
l.h.-side, $Im(\delta)$, generated by the exchange kernel, does not
incorporate the self-energy graphs, where the meson is emitted and
absorbed by the same particle. On the contrary, $\sigma^{inel}$ in
r.h.-side is determined by the inelastic amplitude squared and therefore
contains such contributions. That's why unitarity is violated. It is
restored, if self energy is taken into account.
\begin{figure}[!ht]
\includegraphics[height=.27\textheight]{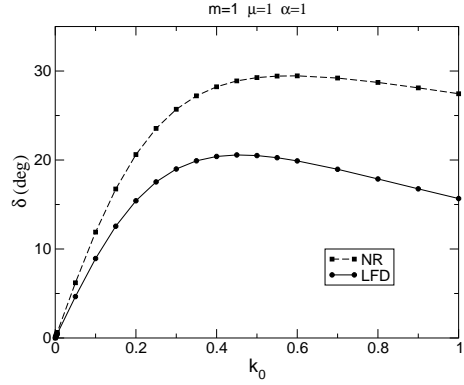}
\caption{Elastic phase shift vs. c.m. momentum $k_0$. Solid and dash
curves are LFD and NR calculations.}\label{fig5}
\end{figure}

Figure \ref{fig5} shows a comparison of NR and LFD calculations of the
phase shift for two scalar bosons exchanging by the mass $\mu=1$. The
conclusion is similar to the bound state case: relativistic and
nonrelativistic results considerably differ from each other even at the
small incident momenta.
\section{Three body systems}
\begin{figure}[!ht]
\includegraphics[height=.26\textheight]{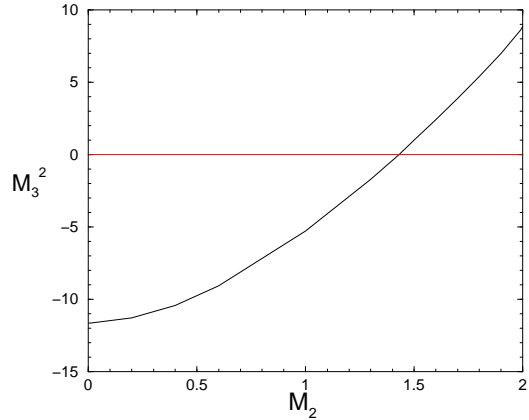}
\caption{Three-boson bound state mass squared  $M_3^2$   versus the
two-body mass $M_{2}$ ( at $m=1$).}\label{fig6}
\end{figure}

Three-boson relativistic system  with zero-range interaction was studied
in \cite{noyes,tobias,ck03}. The input is the two-body bound state mass
$M_2$, the output is the three-body mass $M_3$. As well known,
corresponding NR three-body system is unstable: if, for fixed $M_2$, the
interaction radius $r_0$ tends to 0, then the three-body binding energy
$B_3$ tends to $-\infty$ (Thomas collaps). In relativistic case, it was
found  that  three-body mass $M_3$ is always finite. However, if the
two-body mass $M_2$ approaches to the critical value $M_c=1.43\; m$, then
three-body mass $M_3$ decreases down to zero \cite{ck03}. When $M_2<M_c$,
$M^2_3$ becomes negative (see figure \ref{fig6}).
Therefore, at enough strong interaction, providing $M_2<M_c$, the
three-body system does not exist (there is no any solution with real
$M_3$). This is a relativistic analog of Thomas collaps. For three
fermions the situation is similar, though the critical mass $M_c$ is
smaller.

General spin structure of the LF nucleon wave function composed of three
quarks was studied in \cite{karm98}. Applications to the hadron form
factors were considered in  \cite{bhhk04}.
\section{Many-body contributions}
In general, the number of constituents in a relativistic system is not
fixed. We can consider this system as a few-body one, if few-body sectors
dominate and contribution of higher  Fock sectors is small. The
contributions of two-body and higher Fock sectors to the total norm and
electromagnetic form factor were analyzed \cite{hk04} in the Wick-Cutkosky
model, where  two massive scalar particles interact by the ladder
exchanges of massless scalar particles. Two-body sector contains two
massive particles. Higher sectors contain two massive and $1,2,\ldots$
massless constituents. It was found that two- and three-body sectors
always dominate. Even for maximal value of coupling constant
$\alpha=2\pi$, corresponding to zero bound state mass $M=0$, they
contribute to the norm 64\% and 26\% respectively (90\% in the total).
With decrease of $\alpha$ the two-body contribution increases up to 100\%.
Hence, in this model few-body relativistic system is indeed a good
approximation. This result is non-trivial, since for so strong interaction
one might expect just the opposite relation of few-body and many-body
contributions. However, the cross ladder may correct this result. A few
higher Fock sector contributions to the kernel for massive exchanges were
studied in \cite{sbk98}.
\section{Conclusion}
Relativistic few-body physics covers huge and rich domain of physical
phenomena, including  light nuclei at small distances and hadrons in quark
models. LFD is a very efficient approach to these phenomena. Recent
developments in LFD show good progress in this field.
%
%
\bibliographystyle{aipprocl} 

\end{document}